\documentclass[usenatbib]{basi}
\usepackage[T1]{fontenc}
\usepackage[british]{babel}
\usepackage[varg]{txfonts}
\usepackage{rotating}
\usepackage{dcolumn}
\begin{document}
\title[X-ray grating observations]{Observations of Classical and Recurrent Novae with X-ray Gratings
            \thanks{ By Marina Orio, email: \texttt{orio@astro.wisc.edu} 
}}
\author[M. Orio]%
{M. Orio$^{1,2}$\thanks{email: \texttt{marina.orio@oapd.inaf.it}}\\
$^1$INAF-Padova, vicolo dell' Osservatorio, 5, I35122 Padova, Italy\\
$^2$ Department of Astronomy, University of Wisconsin, 475 N. Charter Str., Madison, WI 53704, USA}
\pubyear{2012}
\volume{00}
\pagerange{\pageref{firstpage}--\pageref{lastpage}}

\date{Received --- ; accepted ---}

\maketitle
\label{firstpage}

\begin{abstract}
 X-ray grating spectra have opened a new window on the nova physics.
 High signal-to-noise  spectra have been obtained
 for 12 novae after the outburst in the last 13 years with the  {\sl Chandra}
 and {\sl XMM-Newton} gratings. They offer the only way to probe the
 temperature, effective gravity and chemical composition of the hydrogen burning white dwarf
 before it turns off. These spectra also
 allow an analysis of the ejecta, which can be photoionized by the 
 hot white dwarf, but more often seem to undergo collisional ionization.
 The long observations required for the gratings have revealed 
 semi-regular and irregular
variability in X-ray flux and spectra. Large short term variability is especially evident 
 in the first weeks after the ejecta have become 
 transparent to the central  supersoft X-ray source.
 Thanks to {\sl Chandra} and {\sl XMM-Newton}, we 
have discovered violent phenomena in the ejecta,
 discrete shell ejection, and clumpy emission regions. As expected, we have 
 also unveiled the white dwarf characteristics. The peak white dwarf
 effective temperature  in the targets of our samples
 varies between $\geq$400,000 K and over a million K, with
 most cases closer to the upper end, although for two novae only
 upper limits around 200,000 K were obtained.
 A combination of results from different X-ray satellites and instruments, including
 {\sl Swift} and {\sl ROSAT}, shows that 
 the shorter is the supersoft X-ray phase, the lower 
 is the white dwarf peak effective temperature, consistently with
 theoretical predictions. The peak temperature is
 also inversely correlated with t$_2$, the time for a decay by 2 mag 
 in optical. 
 I strongly advocate the use of white dwarf atmospheric models
 to obtain a coherent physical picture of the hydrogen burning process
 and of the surrounding ejecta. 
\end{abstract}


\section{Introduction}\label{s:intro}

 The evolutionary track of a post-nova white
 dwarf (WD) has been known since the pioneering work of Gallagher \& Starrfield
 (1976), who discovered that the post-maximum decline in optical light during the outburst of a
 classical or recurrent nova is caused by a shift in the wavelength of maximum energy
 towards shorter wavelengths, during
 a constant bolometric luminosity phase,  close to Eddington luminosity.
 This phase  lasts from weeks to years
 as the white dwarf photsphere shrinks back to pre-outburst radius, while thermonuclear burning
 is still occurring near the surface.

 I will describe the of X-ray grating observations of novae
focusing first of all on a historical perspective.
X-ray observations of {\it quiescent} novae were found interesting early on,
 as an excellent way to study the accretion process in nova close binaries (Becker
\& Marshall 1981). However, the central white dwarf burning hydrogen
 shortly before and after {\it outburst} in its hottest and most
 compact configuration (Shara et al. 1977, Fujimoto 1982),
used to be described as an ``extreme ultraviolet
 source'' not as an X-ray source. The development of
 nova models showed that the white dwarf surface reaches temperatures above 200,000 K,
 and it became clear that not only the ``Wien tail'' of
 the black-body like spectral distribution, but even the peak 
  can fall in the X-ray range (e.g. Prialnik \& Kovetz 1992).
A handful of luminous supersoft X-ray sources (SSS) were discovered with {\it Einstein}, and
 turned out to be white dwarf binaries, although they were not associated
 with nova eruptions (e.g. Long et al. 1981). Three post-outburst
 novae were detected as relatively luminous sources with
 {\it Exosat}, albeit almost without 
 spectral resolution (\"Ogelman et al. 1984, 1987). {\sl ROSAT},
 the new imaging X-ray 
 telescope after {\sl Einstein}, had excellent sensitivity in the softest range,
 so in the late eighties and early nineties
 there was a sudden surge of excitement in the nova community, motivated by the
 possibility of detecting post-outburst novae as extremely luminous and soft 
 sources after the nova ejecta became optically thin to X-rays.  

 Indeed, with {\it ROSAT} the {\it Exosat} source GQ Mus was  observed as a luminous 
SSS more than 9 years after the outburst (\"Ogelman et al. 1993) although it was observed to
 turn off shortly thereafter (Shanley et al. 1995). Furthermore, Nova Cyg 1992 (V1974 Cyg) was
 the brightest X-ray source observed with {\it ROSAT} for a $\approx$ 8 months period starting
 more than a year after the outburst (Krautter et al. 1996, Balman et al. 1998).
 A third SSS {\it ROSAT} nova was detected as far away as the LMC, towards
 which the low interstellar absorption favors the detection
 of soft and luminous X-ray sources (Orio \& Greiner 1999). The SSS phase
 seemed to be mostly short lived: 
 in random, serendipitous observations only 3 out of 38 novae in
 the first 10 post-outburst years were caught in this phase,
 and none out of another 69 other novae observed between 10 and 100 years after the
 eruption (Orio et al. 2001).
 It also became clear that   
 novae in outburst emit a slightly harder, less luminous flux that seemed to be associated
 with the ejected shell.
 This component of the X-ray flux has a different and independent evolution from the WD X-ray source
 (Krautter et al. 1996, Balman et al. 1998, Lloyd et al. 1992, Orio et al. 1996).
 The nebula ejected by a nova can be X-ray luminous for longer than a century
 (like the exceptional
 GK Per, see Balman \& \"Ogelman 1999) although it mostly turns off within 2-3 years
 (Orio et al. 2001, Orio et al. 2009). 

 However,
 at this point in time in the nova scientific
community the WD as SSS was the main object of attention, especially that
of the theoreticians. Model predictions could be tested through these observations.
 The nova and pre-type Ia supernova models predict that the more massive the
 white dwarf, the less mass is accreted until degeneracy is lifted in an outburst.
 This happens because the material 
 is more compressed on the smaller surface of a massive WD, so the pressure needed for the outburst
 is reached sooner. It is thus also reasonable to assume that mass loss ceases
 with less leftover envelope mass on massive white dwarfs, so the 
 SSS duration must be  shorter. There are also other important parameters in
 nova outbursts (mass accretion
 rate, WD initial atmospheric temperature at the
 onset of burning, chemical composition, degree of mixing
 with WD material), but the WD mass is fundamental in determining
the length of the SSS phase.
 
 Certainly not less interesting is the dependence of the highest WD atmospheric effective temperature
T$_{\rm eff}$ reached {\it during} thermonuclear burning on
 the WD mass. The higher  is the mass, the higher should be T$_{\rm eff}$. 
 Even if the {\sl ROSAT} All-Sky-Survey proved that the supersoft X-ray source 
 phase must be mostly short-lived, since no other SSS were detected in other exposures
 of  Galactic novae (Orio et al. 2001), it became known that both symbiotic stars
 and other CV-like systems host at times a hot, hydrogen burning SSS WD for many
 years without apparent outburst, raising great interest for the identification
 of type Ia supernovae progenitors
(e.g. van den Heuvel et al. 1992, Orio et al. 1994).

\section{Why X-ray gratings?}

The so called supersoft X-rays offer a short-lasting, but precious window to observe the 
``naked'' hydrogen burning white dwarf. The WD burns accreted hydrogen in a shell with only a thin
 layer on top, and the atmosphere that becomes extremely hot when the CNO cycle is in full swing.
 The nova
 outburst models indicate temperatures of up a million K. By fitting the continuum and absorption features of
 the hot WD with detailed atmospheric models, in principle we should be able to derive the
 WD effective temperature, radius, and
effective gravity (which indicates the WD mass if the distance to the nova is known).
 Because of the independent correlation of effective temperature and WD mass, we 
 have another way to check whether the derived mass is reasonable.

 The more recent nova models indicate shown that in most CN the WD mass 
 does not grow after each outburst, but rather all accreted mass is ejected and even some WD
 original material is eroded.  However, this may not be true for some recurrent
novae (RN).
 
  With the advent of {\sl Chandra} and
{\sl XMM-Newton} and their X-ray gratings, capable of resolving spectral lines in the soft
X-rays range, the nova community hoped that by observing a number of classical and recurrent novae
 with high spectral resolution, a number of important issues would be solved. They can be summarized as 
follows: 

 1. Are Neon-Oxygen WD common in accreting and hydrogen burning close binaries?
 The ejecta composition of several novae shows overabundant oxygen, neon and
 magnesium (the latter is an enhanced element in NeO WD).
 The final destiny of NeO WD binaries may be an accretion induced collapse, not a type Ia
 supernova. 

 2. Can we distinguish between these three
 possibilities: a) Typical abundances of ``fresh'' CNO ashes (very rich in nitrogen and
 depleted in carbon), b) Freshly accreted material (similar in composition to the envelope of
 the secondary),  or c) chemical composition indicating erosion of the WD core?
 This is crucial to understand whether the WD mass of CN or RN can increase over the
 secular evolution, despite the mass ejection in nova outbursts.

 3. Is the range of peak T$_{\rm eff}$ consistent with the models, and is the temperature
 inversely
proportional to the duration of the SSS phase? This question can be answered thanks to
 correlation with early {\sl ROSAT} observations and especially,
 with the  more recent {\sl Swift}
 monitoring of novae. With short snapshot broad-band observations, even twice a day, the {\sl Swift}
X-ray telescope measures 
 the exact SSS duration (see Schwarz et al. 2011). 
Such monitoring could not be scheduled before {\sl Swift}.

 4. Does the WD return to the ``cold WD'' radius, or is it significant bloated? 

 5. Does mass loss cease completely when the nova fades in optical and UV, or do the X-rays
 indicate a residual wind?  

 These five questions are central for the nova theory, and also for any realistic pre-supernova type Ia
 model from a single degenerate binary system.

\section{A comprehensive summary until 2012 August}

I will not review here the precise characteristics of the X-ray gratings on board
{\sl Chandra} and {\sl XMM-Newton}, well explained in the papers I quote. Suffice
 to say that {\sl XMM-Newton} is equipped with independent X-ray telescopes for
 each instruments, and has two Reflection Grating Spectrometers (RGS) able
 to resolve lines
in the 0.33-2.5 keV energy range, with varying spectral resolution. The effective area peaks around 15 \AA.
{\sl Chandra} is equipped with a low energy transimission grating (LETG) providing 
resolving power (E/$\Delta$E $>$ 1000) at energy 0.07-0.15 keV (80-175 \AA)
and moderate resolving power at higher energies (up to 4.13 keV).
 The HRC camera
has always been used with the LETG in the observations of novae. The HETG (high
 energt transimission gratings) consists of two sets of gratings, each with different period.
 One set, the Medium Energy Grating (MEG), works
 in the 0.4-5 keV range and the High Energy Gratings (HEG) in the 0.8-10 keV energy range.
 The HETG have been used only once for the recurrent nova RS Oph.

Tables 1 and 2 presents a comprehensive list and summary of characteristics of 12 
 novae observed with X-ray gratings. Table 3 presents the data
 on the Large Magellanic Cloud novae (only two yielded
 grating spectra), but {\sl EPIC} data were useful also
 for the other two objects. Before the launch of {\sl Swift} in November
 of 2004, we had to rely on preliminary short observations done with the {\sl Chandra}
 ACIS camera or with XMM-Newton (which uses all detectors - gratings included - simultaneously).
 However, Target of Opportunity (TOO) observations are 
 not practical or easy to schedule with these satellites. 
 Without preliminary measurements, a long
 exposure of nova V1187 Sco was scheduled with the {\sl Chandra} LETG,  but  turned out
to produce no usable results. 
 A TOO mission like {\sl Swift} was needed to proceed to ensure the targets  
 are bright enough for a long ($>$ 3 hours) grating exposure.
Today, after {\sl XMM-Newton} and {\sl Chandra} have worked together 
 with {\sl Swift} for several years, we have learned with precision when the SSS
 became observable 
 in many novae, and an important sample were observed around peak luminosity in 
long exposures with the gratings. There often is
 a correlation between supersoft X-ray phase a Fe [X] forbidden line at 6375 \AA \ in the optical
 spectra (Schwarz et al. 2011),
 but this was not known  13 years ago. In fact, even this optical emission line is
 not an absolute index, since it can also
 be due to shocked ejecta, so it is not a completely reliable
 diagnostic of the SSS observability. In short, having the systematic
 X-ray monitoring has revolutionized the scheduling of grating observations,
 making them really efficient because they were done at or around peak luminosity.
 Since a nova can increase in X-ray luminosity by a factor of 10,000 within
 few days, it is important to observe it at the right time. Fast recurrent novae (RN)
 are very X-ray luminous for only few weeks, as the Tables indicate.

 For the sake of completeness I have also listed  short {\sl XMM-Newton} exposures
 of V598 Pup,
 and of Novae LMC 1995 and 2000,
 although no usable RGS spectra were collected,
 because the aim was to use the {\sl EPIC} cameras, the RGS gratings
would have detected exceptionally bright emission lines if they were emitted.
 So we can rule out at least very prominent and broad
emission lines in this set of novae observed at the given
 post-outburst time.

 The second column in the Tables
reports the outburst date and the third the type (Classical nova, CN, versus RN).
 Because we want to correlate the X-ray data with other wavelengths, especially the optical,
 the fourth column groups
 together three parameters used to measure and evaluate the optical light curve:
 t$_2$ and t$_3$, the time for a decay by 2 and 3 magnitudes respectively, and
 a quality parameter defined by Strope et al. (2010): S for a smooth light curve,  F for a flat-topped
 one (with a plateau), J for ``jitter'', O for the characteristic oscillations of some
 novae, and C for ''cusp'', or light curves with a secondary maximum. We do not have any  
 J-type nova in our sample. t$_2$ and t$3$ were obtained from Strope et al. (2010
 for the Galactic novae, from Subramaiam \& Anupama (2002) for
 the LMC novae, or from AAVSO light curves for more recent novae.
 Columns 6,7,8 and 9 report the date and exposure times with {\sl Chandra}
 and {\sl XMM-Newton}, respectively. For Chandra, an L or an H in front of the exposure date indicate
 either the LETG or the HETG gratings. As we can see, all but one {\sl Chandra} observations
 were done with the LETG, the low energy transmission grating (as opposed to the
 HETG, or high energy transmission gratings). The 10th column indicates the 
 SSS turn off time if it could be measured with {\sl ROSAT} or more often, with {\sl Swift}.
 Finally, we indicate the effective temperature T$_{\rm eff}$  measured by fitting atmospheric models
 in column 11. The measured  T$_{\rm eff}$ in the grating spectra is mostly high, in the 650,000 - one million K
 range. This is mostly because of a selection effect. {\sl Swift} is essentially
a TOO mission, and TOO observations are easier to obtain and to schedule when they are
 done in the same month and for an onvject tha immediately proves to be very luminous (and/or very
 variable). In fact, for the relatively cool and slow N LMC 1995 we have onlyborad band data,
 and the only  ``less hot'' nova with a peak T$_{\rm eff}$ measurement is GQ Mus, observed
 in the {\sl ROSAT} time.   

Quality parameter for the X-ray light curve and one
 for the spectrum are presented in the last two columns.
  For the X-ray light curve (XLC),
 ``f'' indicates that flares were observed, ``p'' and ``sp'' periodic and semi periodic oscillations, and ``o'' an obscuration or sudden dimming of the source.  

   I would like to start the next Session referring to the very last column (12) in the Tables,
 XRS is the attribute given to the X-ray spectrum. A broad classification shows
 that there are three classes of X-ray spectra, those in which the continuum and absorption features of
 the WD are dominant (``b'' for bright), those in which 
broad and strong emission lines in the ejecta are  
 dominant and the continuum is hardly measurable with the gratings (``e'' for emission), and the
 hybrid ones where the two components appear equally important. If no parameter is listed
 in this column, no usable grating spectrum was obtained. In some cases, the 
same nova transitions
 between these states during the outburst (e.g. RS Oph, N LMC 2009). 

\begin{figure}
\centerline{\includegraphics[width=7.1cm]{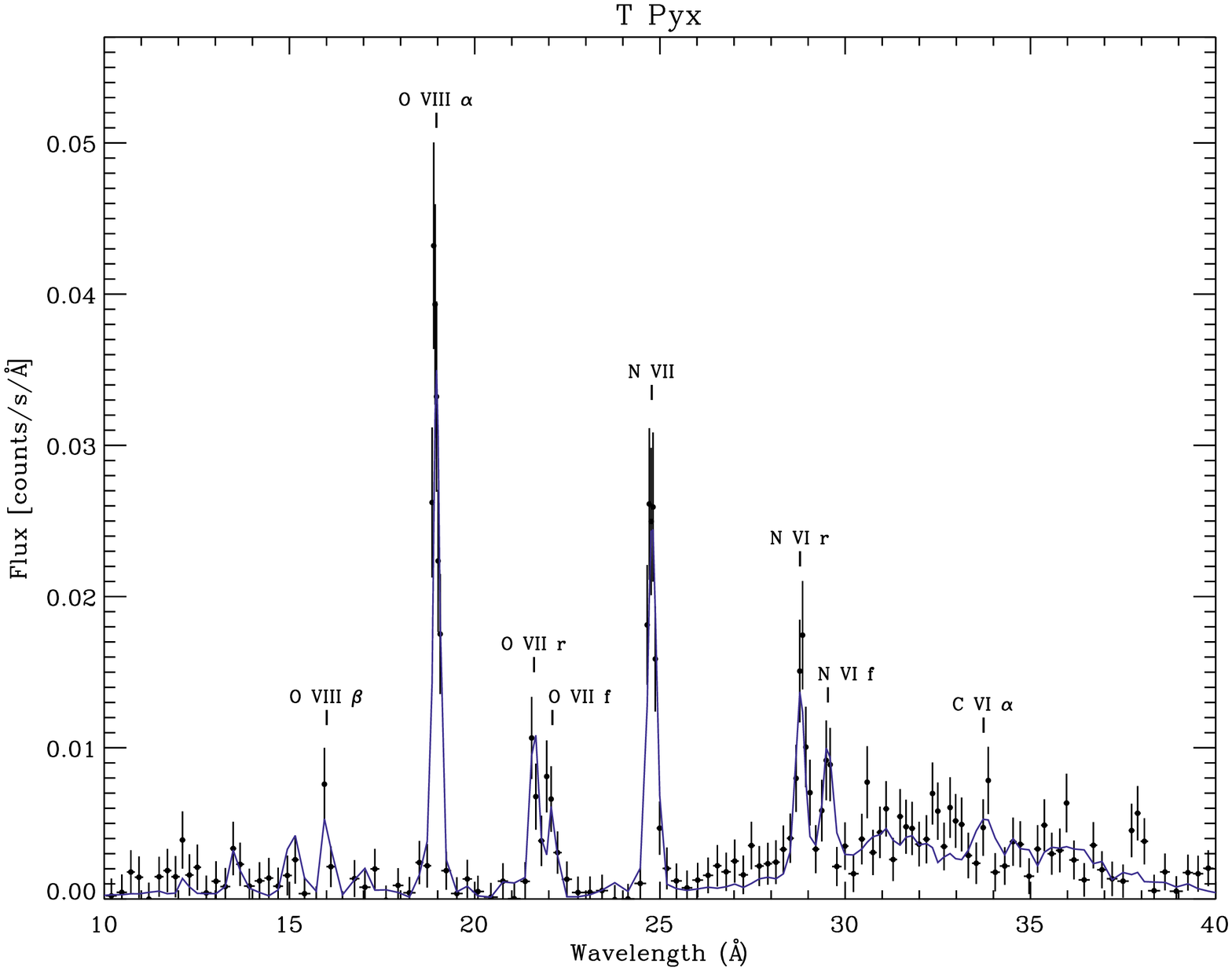} 
  \includegraphics[width=6.6cm]{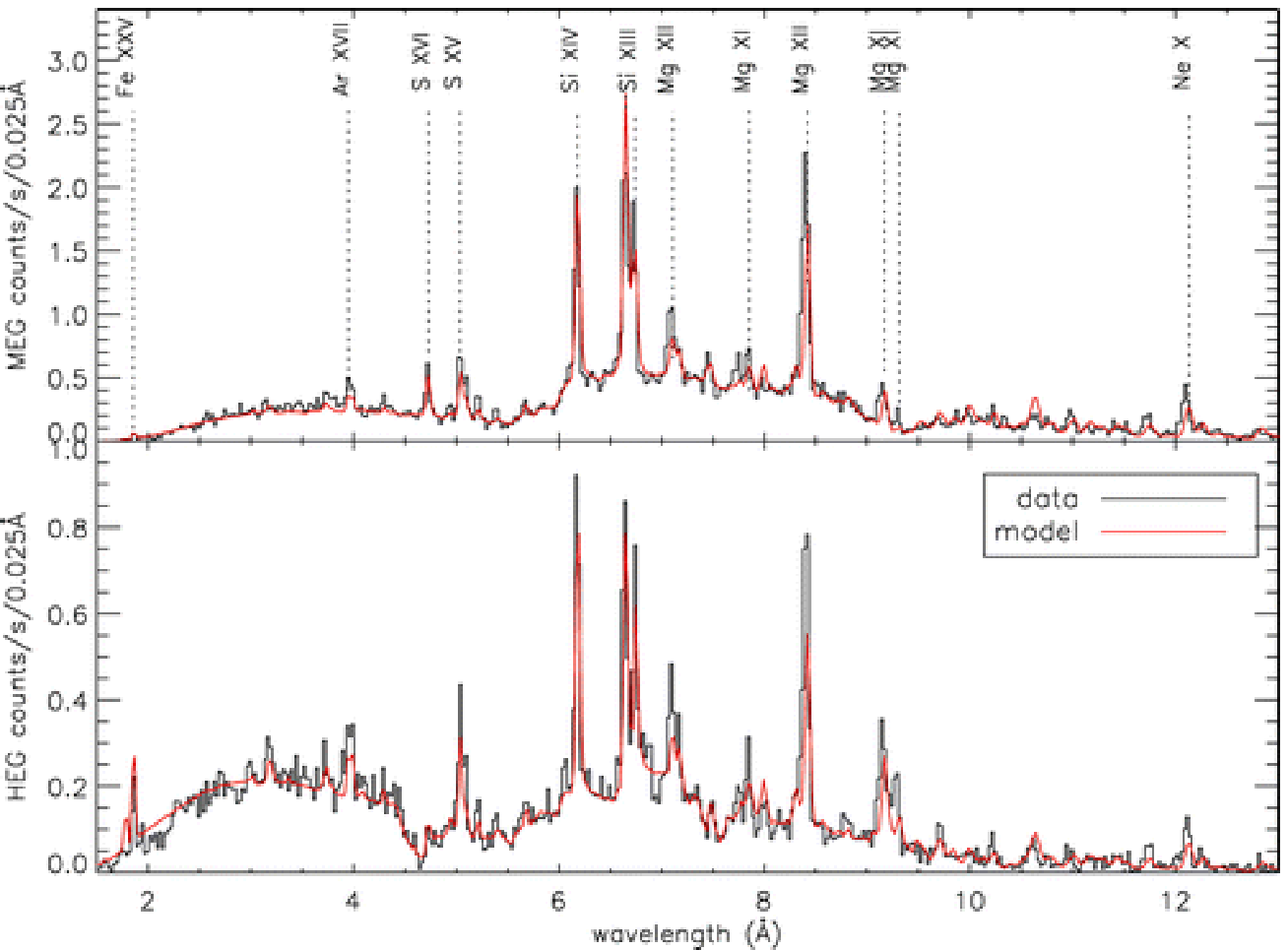}}
\caption{\footnotesize{Novae spectra dominated by emission-lines ({\it e} type
 in the Tables) : LETG spectrum of T Pyx in October
 of 2011 (2a, left) and HEG and MEG spectra of RS Oph in February of 2006 (2b, right), with
 collisional ionization models.
 RS Oph was observed with the LETG in April of the same  year and a luminous atmosphere
was detected, with only few residual emission lines in the soft range.
}}
\end{figure}

\section{The unexpected emission lines}

Although the X-ray emission in the ejecta seemed to be known as a fact by the time 
 the X-ray grating observations started in May of 1999, the spectra obtained for the first two
 novae, V382 Vel and V1494 Aql, of the ``e'' type, were quite surprising.
 It was suddenly clear that the so-defined ``thermal bremsstrahlung''
 continuum in other novae (e.g. Orio et al. 1995, Balman et al. 1998) was
 not a continuum, but there were  many unresolved strong and broad emission lines.

 Orio et al. had already obtained broad band spectra of V382 Vel with {\sl BeppoSAX},
 (published in 2002), probably shortly after
the SSS maximum. Although the nova was a luminous SSS, it was obvious that the
 continuum departed so much from atmospheric models that other components were present
 and even emitted conspicuous flux. In two and a half months intervening until
 the first {\sl Chandra} LETG grating observation could be scheduled, this second
 ``component'' was dominating. The WD was much less luminous as an SSS, having started to
 turn off, with T$_{\rm eff} <$ 300,000 K. It displayed 
 many emission lines, due to hydrogen-like and helium-like transitions of Si, Mg, Ne, O, N and C 
in the 6-50 \AA \ range. Iron lines were completely missing.
 All the lines had broad profiles with full width at half maximum
 corresponding to about 2900 km s$^{-1}$, and a few had a complex structure, 
 to be fitted with at least two Gaussians (Ness et al. 2005).   
 Altogether, all this is rather typical of other novae, as well (see Fig. 1).

  In observations of V1494 Aql seven months later, the emission lines were very dominant
 again,
 although there was a non-negligible continuum level. A complete analysis for this nova
 appeared only later, in a paper by Rohrbach et al. in 2009. Emission lines were observed
 this time in the 16-34 \AA \ range, and  although the authors could fit a comprehensive model,
 they noted that at least the O VII Lyman-alpha line can only be explained with
 collisional ionization, because an upper limit to the
 temperature of the central WD is around 
 200,000 K. The authors favored however a combination of collisional and photo-ionization
 to explain the rest of the spectrum. 

  It was not until the 2003 observations
 of V4743 Sgr that the luminous SSS WD could really be studied in great detail.
 For this nova however,
 a sudden obscuration of the WD during the first observation revealed a low level emission
 line spectrum that was superimposed on the WD atmosphere. Emission lines thus always 
 exist in the X-range for novae. Some grating spectra (HV Cet, V5156 Sgr, U Sco)
that we classify as ``be''
 show a measurable continuum and at the same time very strong emission lines, making the
 fit with  composite models accounting for both the WD atmosphere 
and the ejecta quite challenging.
 For U Sco, collisional ionization in a dense and clumpy
 medium  has been invoked to explain the line ratios
 of the observed triplets (Orio et al. 2012a). The emission lines seem to be produced in 
 condensations of very dense material.

  For some
 objects, fortunately, the emission lines are less strong and the
 continuum more luminous, so the WD can be better studied. T Pyx, observed
 in an eruption at the end of 2011 (see Fig. 1, left panel), is 
another nova for which strong emission lines were detected,
 and only an upper limit can be given for the T$_{\rm e}$ of the central
 source, about 200,000 K (Tofflemire et al. 2012).
 There never was very luminous period during the complete {\sl Swift} 
coverage of T Pyx, and  a large part of the
 SSS flux seems to have always been in line emission.
 The T Pyx gratings spectra are consistent with the ``low'' effective temperature of
 a small mass WD, and (mostly) with a collisional
 ionization mechanism of emission line formation.
 Yet,  T Pyx is a RN, where high mass and effective temperature  are expected,  since 
 the envelope pressure for an outburst is frequently reached.
  
 The most surprising emission line spectrum ever observed is the one of RS Oph
 shortly after the outburst, when also the {\sl Chandra} HETG gratings    
 were used (Fig. 1, right panel), in addition to the RGS. For this nova 
 the emission lines were in a very
 hard range in the first month after the outburst. It was already
 known that the nova ejecta collide with the circumstellar material
 of the red giant wind in the RS Oph binary
 system, yet this hard spectrum, certainly  a proof of violent shocks, 
 was surprising. This spectrum evolved and disappeared quickly. 
Four months later, after the SSS had became
 bright and turned off, the RS Oph spectrum was similar to the one of T Pyx on the left. 

\begin{figure}
\begin{center}
\begin{tabular}{p{6cm}cp{6cm}}
\raisebox{-\height}{\includegraphics[width=7cm]{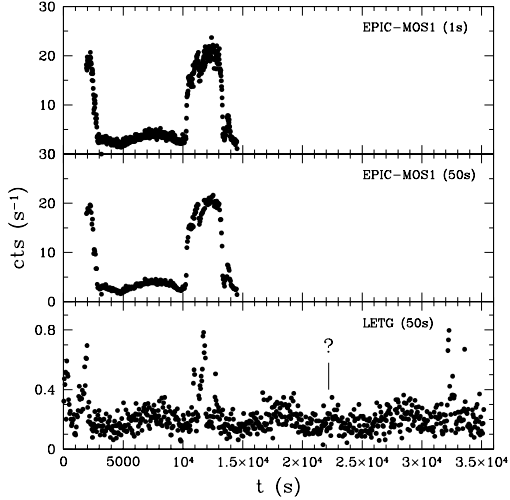}} & \quad &
\caption{The upper and middle panel show the EPIC-MOS-1 light curve observed for V5116 Sgr 
 plotted with bins of 1 and 50 s, respectively. This observation was
 described by Sala et al. (2008). The bottom panel shows the 
{\sl Chandra} + HRC-S detector and LETG grating light curve, done
 5 and a half months later. This observation
 was longer and covered four orbital cycles. In three out of four orbital cycles the flaring
 behavior was clearly observed. Note that the amplitude of the
 flare, however, significantly decreased in the {s\l Chandra} observations (plot from Bianchini \& Orio,
 2012b, in preparation).}
\end{tabular}
\end{center}
\end{figure}

 \section{Wild variability after the outburst}
 Before proceeding with the analysis of what has been learn thanks
 to the SSS spectra, it is important to review the short term variability of
 the post-nova SSS.
 The exceptional variability of post-novae WD was apparent already after the second
X-ray  observation done with a grating. Exposure times with previous satellites were
 of the order of 1-3 hours for pointed observations with CCD-type
 detectors, and exposures were only  a few
 minutes long during the  {\sl ROSAT} All-Sky-Survey.
 The need to expose for longer than previously done with broad band detectors,
 in order to obtain a reasonable signal-to-noise with the
 X-ray gratings, implies also that variability phenomena are easier to detect.
 Generally, the variability seems more common in the first weeks after the SSS turns on,
 and it is quite clear that it has several different root causes. The observed
 phenomena are 
 periodic, quasi-periodic and periodic, on  time scales from tens of seconds to many
 hours. I will attempt to classify these phenomena as follows:

  1) {\it Orbital variability}. Modulations of the soft X-ray flux
 with the orbital period were already observed with {\sl ROSAT} for GQ Mus  
 (Kahabka 1996) and possibly 
 N LMC 1995 (Orio et al. 2003). they have been more recently measured
in X-ray gratings exposures for
V4743 Sgr (Leibowitz et al. 2006), U Sco (Ness et al.
 2012, Orio et al. 2012a) and T Pyx (Tofflemire et al. 2012). The modulation
 is associated with the central source for U Sco (the emission
 lines were unvaried), but there is also modulation of
 the emission lines in T Pyx.  Another SSS nova, HV Cet, was observed with {\sl Swift}
 to have a 1.77 days orbital modulation (Beardmore et al. 2010; this orbital period
 was too long to be observed with the  {\sl Chandra} exposure).

  2) {\it Periodic or quasi-periodic variability} with a period (or quasi-period) varying
 between tens of seconds and tens of minutes. The fastest oscillations were observed for
 RS Oph and KT Eri ($\simeq$35 s in both cases). For RS Oph the period was not stable and changed
 from day to day, but the last long {\sl Chandra} observation and {\sl Swift}
 observations during the plateau of the SSS phase show that the period became stable
 over several days before disappearing (Leibowitz et al. 2012, unpublished). Leibowitz
 et al. (a project in which the author is involved) are proposing that this is the rotation
 period of the WD in RS Oph, and that the initially apparently 
``unstable'' period is due to the collimated
 outflows (observed at radio wavelength, Eyres et al. 2009): the jet may have been 
 precessing as it left the WD polar caps' surface and was 
 hot, still emitting super-soft X-rays. This precession scenario fits the data
 quite well. Some of our colleagues
 have criticized this explanation as too speculative, but other explanations are 
difficult to devise, unless the oscillation is still associated with the WD spin.
 Indeed, theoretical predictions by Yoon \& Langer (e.g 2005 paper) foresee
 a rapidly rotating, with periods of order of a minute for a WD spun up by high accretion rate (as 
 is the case for KT Eri and RS Oph).

 A much longer oscillation lasting  $\simeq$41.7 min was observed in V1494 Aql
 with {\sl Chandra}, several periods of tens of minutes were measured for V4743 Sgr, and a 37.2 min 
 period for  V2491 Cyg (Ness et al. 2011). These may be better explained with non-radial
 g-mode pulsations of the hot WD. However, there is no theoretical model for these pulsations.
 They may be associated with iron ionization, but not with ionization of
 other elements as in ``cold'', non-accreting WD.   
 A number of side-frequencies observed in V1494 Aql and V4743 Sgr seems to support
 this explanation. Things are a little more complex however, since one of the periods
 of V4743 Sgr is observed until and during quiescence and at optical wavelengths, and
 seems to be the spin period of the WD (Leibowitz et al. 2006, Dobrotka \& Ness 2010).
 V1494 Cyg, V4743 Sgr, and V2491 Cyg are all thought to be intermediate polars (IP) also
 for other, independent reasons.  

 On the other hand, U Sco is not an IP and it and a semi-regular oscillation with
 a period of about half an hour in the first {\sl XMM-Newton} exposure is explained
 by Ness et al. (2012) with dense clumps of material that fall in the newly re-estabilished
 accretion disk. We have suggested that this could be the disk rotation period (Orio et al.
 2012a). 
 
 3) {\it A sudden dimming or obscuration of the WD}. The most dramatic case is V4743
 Sgr in the first {\sl Chandra} observation, in which after several hours of observations
 of the SSS, the central source suddenly disappeared within few hours.  
 Two weeks later, when {\sl XMM-Newton} pointed at the nova, the SSS was luminous again
 and there was not any  new obscuration in this and several successive observations.
 This phenomenon may be explained with a sudden shell of 
 material that leaves the WD surface and is optically thick to supersoft X-rays.
 If this is the correct interpretation, the nova outflow is not always smooth and continuous,
 but rather episodic.

4) {\it A large and sudden flare} lasting for about 1000 s was observed for V1494 Aql (Drake
 et al. 2003), A very similar, {\it periodic} flare almost at each orbital period was observed for
 V5116 Sgr with {\sl XMM-Newton)} (Sala et al. 2008) and months later with {\sl Chandra}
 by the PI, albeit with much
 lower amplitude (see Fig. 2). Although Sala et al. have proposed an explanation in terms of a warped,
 truncated disk, it is difficult to imagine how such a disk could have been stable
 for months, and why the rise occurs within only few minutes.   
 I suggest that the root cause may be a non completely thermalized atmosphere in 
 a magnetic system, an intermediate polar or polar nova
 where the polar caps were much hotter than the rest of the surface at the time of the
outburst. This does not rule out that the thermonuclear flash propagated evenly
 in the bottom layers of the envelope. 
 
\begin{sidewaystable}
\caption{\small Galactic novae in outburst observed with X-ray gratings, including
 XLC=observed optical maximum date, the type (classical or recurrent), optical light curve parameters t$_2$, t$_3$ and
 Strope quality parameter (see text), time of Chandra or XMM-Newton observation with
 exposure time in kiloseconds, time  for the SSS turn-off, WD effective temperature, 
 XLC or X-ray variability parameter (f for flare, o for occultation, p for periodic, sp for
 semi-periodic, po=periodic with orbital period),
 and XRS, type of X-ray spectrum (e=emission lines, b= dominated by very bright white
dwarf, be=hybrid, no=no results. The atmospheric fits were done by the author (see also references).} 
\begin{center}
\begin{tabular}{llllllllllll}
\hline
 1   &  2    &  3   & 4 &      5          &      6     &   7  &   8 & 9  & 10 & 11 & 12 \\
Nova & Max.  & Type & LC & Chandra & exp. & XMM-Newton & exp. & t.off & T & XLC & XRS \\
     &  (date)    &      &    & (date)        & ksec & (date)  & ksec & months & 10$^3$ K & & \\
\hline
V382 Vel (1,2) & 5-22-99 & CN & 6,13,S & L 3-2-01 & 24.5   & & &  7$\pm$1      & $\geq$400 & o & e \\
V1494 Aql (3,4) & 12-3-99 & CN & 8,16,O & L 9-28-00  & 8.1 & & & 19.17$\pm$7.05 & & f & e \\  
          &         &    &        & L 10-1-01  & 18.2  & & & & & & e \\
          &         &    &        & L 11-28-01 & 25.8  & & & & & & no \\
V4743 Sgr(5,6,7) & 9-20-02 & CN & 6,12,S & L 3-19-03  & 25 &        &     & 20.5$\pm$3.5 & 740$\pm$70 & o,p & b \\ 
          &         &    &        &            &    & 4-4-03 & 35.6 &         & & & b \\
          &         &    &        & L 3-24-06  & 10 &      &      &         &  & sp & b \\
          &         &    &        &           &     & 4-7-06 & 18.6 &        &   & sp & b \\
          &         &    &        & L 7-18-03 & 12  &        &      &        &   & sp   & b \\
          &         &    &        & L 9-25-03 & 12  &        &      &        &   & sp   & b \\
          &         &    &        &           &     & 9-30-03 & 22.5&        &   & sp & b \\
          &         &    &        & L 2-28-04 & 10  &         &     &        &   & sp & b \\
V5116 Sgr (8,9) & 7-4-05  & CN & 12,26,S  &          &       & 3-5-07 & 13 & 34.5$\pm$3.5 & 700$\pm$100 & f,p & b \\
          &         &    &          & L 8-24-07 & 12 &        &         &  & & f,p & b \\ 
RS Oph (10,11)  & 2-12-06 & RN & 7,20,S   & H 2-26-06 & 10  & 2-26-06 & 23.8 & 3.25$\pm$0.25 & 800$\pm$50 & & e \\ 
          &         &    &         &           &      & 3-10-06  & 11   & &   & f,sp & be \\
          &         &    &         & L 3-24-06 & 10   &          &      & &   & o, sp   & b \\
          &         &    &         &           &      & 4-7-06   & 18.6 & &   & sp & b \\
          &         &    &         & L 4-20-06 & 9    &          &      & &   & p & b \\
          &         &    &         & L 6-4-06  & 20   &          &      & &   & & e \\
          &         &    &         &           &      & 9-7-06   & 10   & &   & & e \\
          &         &    &         & L 9-7-06  & 40   &          &      & &   & & e  \\
          &         &    &         & L 10-9-06 & 40   &          &      & &   & & e \\
          &         &    &         &           &      & 10-9-06 & 48.7 & &   &  \\
V598 Pup (12)  & 6-4-07 & CN &         &           &       & 10-29-07 & 5.1 & & &  \\
\hline
\end{tabular}
\end{center}
\end{sidewaystable}
\begin{sidewaystable}
\caption{\small Continuation of Table 1}
\begin{center}
\begin{tabular}{llllllllllll}
\hline
 1   &  2    &  3   & 4 &      5          &      6     &   7  &   8 & 9  & 10 & 11 & 12 \\
Nova & Max.  & Type & LC & Chandra & exp. & XMM-Newton & exp. & t.off & T & XLC & XRS \\
     &  (date)    &      &    & (date)        & ksec & (date)  & ksec & months & 10$^3$ K & & \\
\hline
V2491 (13) Cyg & 4-10-08 & RN? & 4,16,C &           &      & 5-20-08  & 39.3   & 1.47 & 1000$\pm$100 & o,p  & b \\ 
          &         &     &        &           &      & 5-30-08  & 30 &  &   & & b \\
HV Cet (14,15) & 10-7-08 & CN &         &  L 12-18-08 & 35   &         &     & 3.5$\pm$1   & 750$\pm$100 & & be \\ 
KT Eri  & 11-14-09      & CN & 15,28,S &  01-23-10 & 15 &   &  & 9.5  & 650$\pm$150 & sp & b \\   
        &               &    &         &  01-31-10 & 5.2 &  &  &      &             &    &   \\
        &               &    &         &  02-26-10 & 5.2 &  &  &      &             &    &   \\
        &               &    &         &  04-21-10 & 5.2 &  &  &      &             &    &   \\
U Sco (16.17) & 1-28-10 & RN & 1.2,2.6,S & L 2-14-10 & 23 &          &  7  & 1.67 & 950$\pm$100 & & be  \\
           &         &    &        &           &      & 2-19-10   & 64 &       &     & po,sp & be \\
           &         &    &        &           &      & 3-13-10   & 63 &       &     & po  & be \\
T Pyx (18) & 4-14-11 & RN & 32,62,P & L 3-11-11   & 40   &           &    &       &      & p & e \\
           &         &    &        &           &      & 11-28-11 & 30  &  &           & p & e \\
\hline
\end{tabular}
\end{center}
{\it {\small 1) Orio et al. 2002, 2) Ness et al. 2005, 3) Drake et al. 2003, 4) Rohrbach et al. 2009,
5) Ness et al. 2003, 6) Leibowitz et al. 2006, 7) Rauch et al. 2010, 8) Sala et al. 2008, 
9) Nelson \& Orio 2009, 10) Nelson et al. 2008, 11) Ness et al. 2011a,  12) Page et al. 2009 
13) Ness et al. 2011b, 14) Sala et al. 2006, 15) Nelson \& Orio 2007, 16) Orio et a;. 2012a, 17) Ness
 et al. 2012, 18) Tofflemire et al. 2012   
}}
\end{sidewaystable}
\begin{figure}
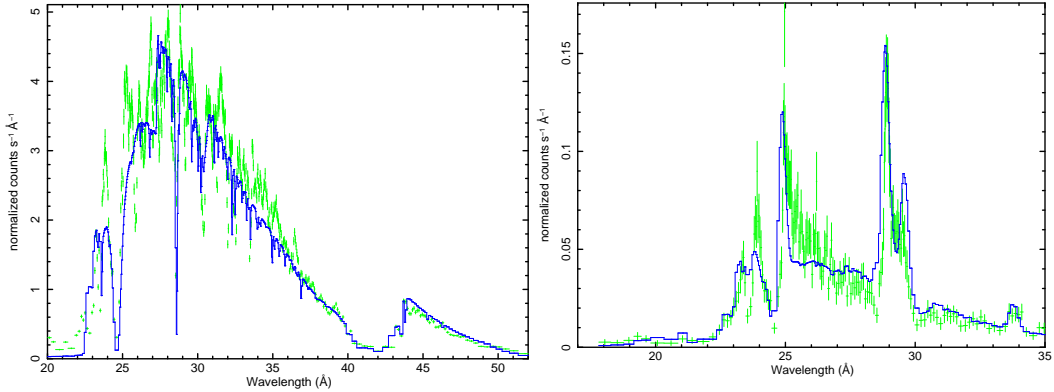

\centerline{\includegraphics[angle=-90,width=6.9cm]{PlotV470303.ps}
  \includegraphics[angle=-90,width=6.9cm]{usco.ps}}
\caption{\footnotesize {
{\sl Chandra} LETG 
 grating spectra of novae in the luminous WD phase: V4743 Sgr (``b'' type) in March of 2003
(left panel), and U Sco (``be'' type) in February of 2010 (right panel).
The U Sco spectrum is in green, the fit with a model in blue. We show an atmospheric model at
 700,000 K for V 4743 Sgr, and an atmospheric model+collisional ionization for U Sco. Note the
 different strength of the emission lines relative to the continuum in the two novae, and that
 some emission lines are still unidentified.}}
\end{figure}
\begin{figure}
\centerline{\includegraphics[width=8cm,angle=-90]{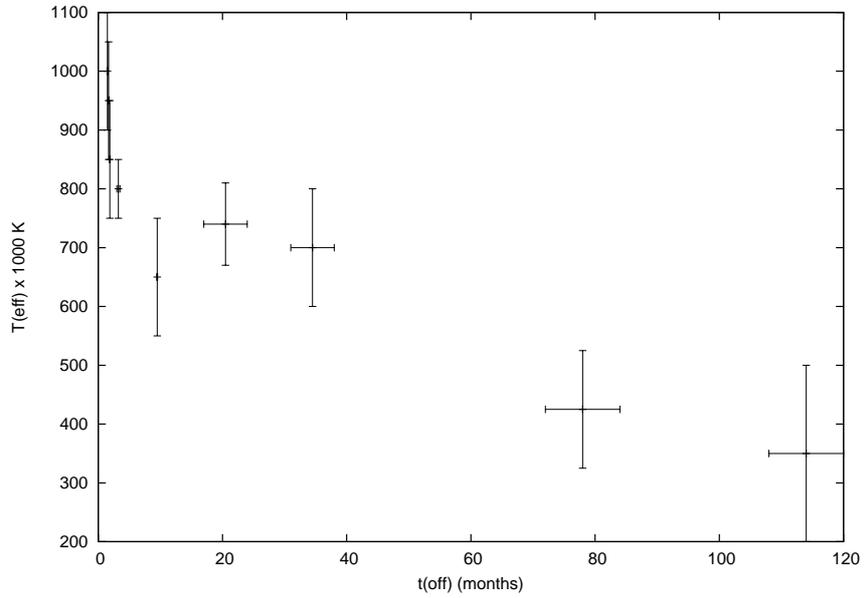}}
\caption{\footnotesize{Turn off time versus peak effective temperature {\it measured with atmospheric
 models}. The two last data points on the right indicate measurements made with broad band detectors.}
}
\end{figure}
\begin{figure}
\centerline{\includegraphics[width=8cm,angle=-90]{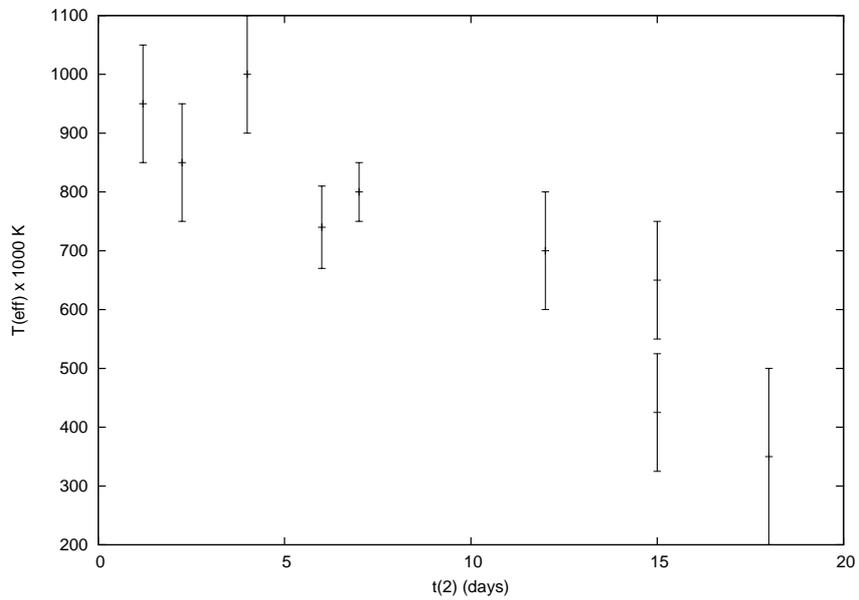}}
\caption{\footnotesize{Optically measured t$_2$ versus peak effective temperature {\it measured with atmospheric
 models}. The two last data points on the right indicate measurements made with broad band detectors.}
}
\end{figure}

\begin{sidewaystable}
\caption{\small Magellanic Clouds novae observed with X-ray gratings, including
OLC observed optical maximum date,
 type (classical or recurrent), optical light curve parameters t$_2$, t$_3$ and Strope
 quality parameter (see text), time of Chandra or XMM-Newton observation with
 exposure time in kiloseconds, time for the SSS turn-off, WD effective temperature,
 XLC or X-ray variability parameter (f for flare, o for occultation, p for periodic, sp for
 semi-periodic, po for periodic with orbital period), and XLS or type of X-ray spectrum.}
\begin{center}
\begin{tabular}{llllllllllll}
\hline
 1   &  2    &  3   & 4  &   5     &  6  &   7         & 8    &  9    & 10 & 11 & 12 \\
Nova & Max.  & Type & LC & Chandra & exp. & XMM-Newton & exp. & t.off & T & XLC & XRS \\
     &  date    &      &    &   date & ksec & date  & ksec & months & 10$^3$ K & \\
\hline
& & & & & & & \\
N LMC 1995 (19,20) &  & CN    & 15,23 &         &           &      &             &  78$\pm$6 & 425$\pm$100 & p? &  \\
N LMC 2000 (21) &  & CN    & (),17 &         &           &      &             &  9.8$\pm$4 & &     &  \\ 
N LMC 2009 & 2-5-09 & RN &         &           &      & 5-6p-09    & 37.6 & 10 & & & be \\
           &        &    &         &           &      & 7-20-09   & 58   &         & & & b \\
           &        &    &         &           &      & 8-20-09   & 32   &         & & & b \\
           &        &    &         &           &      & 9-23-09   & 51   &         & & & b \\
N LMC 2012 & 3-26-12 & RN & 2.25, 4.15 & L 4-26-12 &  20.2 &   &      & 1.87    & 850$\pm$80   & & b \\
& & & & & & & \\
\hline
\end{tabular}
\end{center}
{\it {\small 19) Orio et al. 1998, 20) Orio et al. 2005, 21) Greiner et al. 2004}}
\end{sidewaystable}
\section {The WD spectrum}

 In most novae, we have observed a very strong continuum and at least
resonance absorption lines of the two highest ionization stages of C, N, and O.
 Fitting and interpreting the WD spectrum is clearly hindered by the superimposed
 emission lines, which in some cases make it difficult to trace
 the continuum shape. The ``wild variability'' described above is another
 problem. Good quality
time resolved spectra over the variability time scales can 
 only be obtained at times, especially of repeated cycles were observed. The
 $\simeq$35 s oscillations of RS Oph and KT Eri are too short to extract a spectrum,
 at the peak
 of oscillations, for instance. For U Sco, phase resolved spectra could be extracted, demonstrating 
 significant variation of the emission lines over
 the orbital period or along the oscillations of
 the second observations (Ness et al. 2012, Orio et al. 2012).
 Ideally, long grating exposures are desirable, especially at the onset of the SSS phase,
 when it is almost certain that some type of variability will be observed.

 Another interesting complication is the possibility, especially at high inclination,
 that the WD is actually not observed because the accretion
 disk has already been rebuilt. In U Sco, for instance,
 optical observations indicate a disk already two weeks
 after the outburst. However, a Thomson scattering corona makes a 
 ``reflected'' continuum observable. In these cases, like for 
 Cal 87 (Orio et al.  2003), the total bolometric luminosity phase is only a small
 fraction of the near Eddington luminosity of objects like RS Oph or V4743 Sgr,
 and is clearly not consistent with T$_{\rm eff}$, if this can be
derived, and with the radius obtained from the effective gravity g.
 For U Sco, only about 30\% of the WD continuum was observed to be eclipsed
 by the secondary. The rest of the SSS flux comes from a more extended area than the orbital separation.
Initially the broad absorption features were not embedded in the extremely
 broad emission lines. It was concluded by two
 different groups (Ness et al. 2012, Orio et al. 2012)
that that about 70\% of the flux is emitted
 by a surface with a large
radius, a Thomson scattering corona. Although Thomson scattering is not wavelength dependent,
 it may smear the absorption features, depending on the geometry of the system.
 Model fitting may thus become more difficult and less reliable.

  Given the much higher X-ray flux measured
 for the other novae in the SSS phase, it seems that cases
 like Cal 87 or U Sco are relatively rare. However, it has been argued that
 the main complication in understanding nova spectra is the large blue shift of the
 absorption features observed in some spectra, most notably V4743 Sgr (Rauch et al.
 2010), V2491 Cyg (Ness et al. 2011) and U Sco (Orio et al. 2012).
 The group of Ness et al. (e.g., 2011 paper) 
 has repeatedly argued that no hydrostatic atmospheric models can give reliable
 results because a wind is still flowing from the
 the central SSS. Their reasoning is that this SSS cannot be the WD itself.
 My objection is that: a) The ``moving lines'' were observed to be extremely
 similar (except for the blue-shift)
to other cases in which the absorption lines were instead observed at zero velocity,
 e.g. RS Oph. This seems to demonstrate that the wind
 must really be at the base of the atmosphere, still reflecting its characteristics,
and b) Both blackbody or atmospheric model yields such high temperature
 that the SSS must be extremely compact, or else it would be much above Eddington luminosity.
 Not surprisingly, atmospheric models often yield a log(g)=9 and
 masses above 1.2 M$_\odot$.

   It turns out that the observed novae were mostly chosen as targets because {\it Swift}
 had discovered the SSS phase. This is extremely helpful, but
 it also carries a strong bias towards fast novae. Not even {\sl Swift} will keep
 on monitoring a nova every week, if not every day, for many years. A nova on a 
 non-massive WD, whose ejecta are also more massive,
becomes optically thin to the SSS after years, possibly only shortly
 before the SSS turns off, and will be missed. Not surprisingly, we
 mostly derived very high T$_{\rm eff}$ novae in our sample in Tables 1,2 and 3,
 which thus is clearly biased.

 Atmospheric models for the hottest, hydrogen burning WD have long been known to
 yield fits with higher T$_{\rm eff}$ and lower luminosity than blackbodies, 
 because a realistic atmosphere always departs from a simple blackbody approximation
 (Heise et al. 1994). After initial work by Hartmann (e.g. Orio et al.
 2003 and references therein),
 remarkably complex non-LTE models have been  calculated by Lanz et al. (2005)
 and by Rauch et al. (2010). Rauch has made his models publicly available,
 and even Ness et al. (2011), despite their
 objections to a non-dynamical mode, remark that the fit to the V2491 Cyg spectrum
 is really quite good. I argue that non-LTE model atmospheres
 give a physically significant qualitative agreement with the observed spectra even when a velocity field
 is present, and should be used.
 Rauch et al. employ plane-parallel, static models calculated with the
TMAP10 code (Werner et al. 2003). Opacities of elements H, He, C, N, O, Ne, Mg, Si, S,
and Ca-Ni are included.  The grid of available models has been studied for V4743 Sgr
 and U Sco, and in principle new models should be developed for every
 nova to really obtain a good fit, even if it is a heavy ``numerical task''.

 A recent attempt to explain the V2491 Cyg absorption spectrum has been done by
 Pinto et al. (2012). The authors adopt a blackbody for the continuum, and assume
 that its spectrum is absorbed by three different ionized absorbing shells. The
 absorption includes dust and is different for every shell. Even if it is worth
 exploring how mass ejection may have been discrete,
 and it is important also to implement realistic opacities,
 it seems almost a wasted
 effort because the extremely luminous central source is modeled with a blackbody,
 neglecting the deep and strong absorption features of the central source itself.
 Whatever absorption lines may be produced in the ejecta, must be superimposed
 on the WD ones.
Altogether, Pinto et al. (2012) present
 an interesting model, but there are many ad hoc assumption. 
A weak point is this: Why are no emission features observed in V2491 Cyg
 that correspond to the absorption ones?
 Red-shifted emission causing a P Cyg profile is observed in the optical
 spectra of novae and lasts always longer than the absorption,
 coming from a larger volume; in other novae X-ray emission
 lines attributed to collisional ionization are observed, but not in V2491 Cyg.
 The ejecta of V2491 had reached a distance 
 that was overwhelmingly large compared with the WD radius when the observations were
 made, more than a month after the eruption - it seems difficult not to observe the
 corresponding emission. The authors hypothesize a possible large asymmetry in the shell.
 In short, in addition to not giving a better spectral fit than model atmospheres, the 
 model by Pinto et al. (2012) is not self consistent yet and makes many
 assumptions. We cannot rule out that some absorption features may be formed in 
 a thin shell in the ejecta as in the model, but certainly they would not appear
 as broad as in U Sco, for instance. 

 Even if Lanz et al. (2005) note that the perfect atmospheric model for a hot, hydrogen
 burning WD cannot be calculated yet for deficiencies in
 the atomic data, I already see a remarkable strength of the model atmospheres: their 
 agreement with the observed spectra, and their test  
of the theoretical models. This test can be seen in
 Fig. 4 and 5. The correlation of WD T$_{\rm eff}$ with SSS turn-off time and with
 t$_2$, a parameter obtained form the nova speed class and is also correlated with
 m$_{\rm WD}$, are quite good. These plots clearly demonstrate the theory's predictions,
 that the fastest novae are also the most massive, if we assume that T$_{\rm eff}$ is a proxy
 for the WD mass M$_{\rm WD}$. Of course, a spread is expected  in this
 relationship, due to the other important parameters in determining
 the post-outburst evolution: mass transfer rate $\dot m$, initial T$_{\rm eff}$ of the WD and
 its chemical composition, with the first being ofyten the most dominant. 
 Note that in the figures, I added the {\sl ROSAT} source GQ Mus at the low temperature end (in 1993 it as fitted
 with an early, less complete model atmosphere, courtesy of J. Mac Donald). I would like to
 point out that here the definition of the SSS and its turn-off is different from  Schwarz's et al. (2011).
 My definition is  based on spectral fits and  T$_{\rm eff}$>200,000 K; it is  more rigorous and physics based
 than definition based on hardness ratio. Schwarz et al. include many more low luminosity
 object, where there are no statistics or spectral resolution to rule out that we
 are observing an emission line spectrum and no SSS at all. The luminosity of some of these novae
 appears too low to be due to the central WD.

\section{Conclusions: What are we learning?}

 Cecilia Payne Gaposhkin presented and reviewed optical light curves and spectra  
 of classical and recurrent novae in her book ``The Galactic Novae'', which is still
 fascinating to read for the nova expert. X-ray light curves and spectra are revealing
 a similar wealth of complex phenomena, and it is taking years to analyse them. For each
 X-ray grating observations, more than one paper has appeared, and mostly these
 paper have taken a while to complete, they were not immediately published.
 
 The emission lines spectra of novae are demonstrating the complex physics of the mass
 outflows. Collisional ionization seems common in nova shells, and the ejecta are far
 from being homogeneous, they are rather clumpy. Mass seems be lost at times in ``parcels''
 or discrete shells. There are still several unidentified emission lines.
Most of them are listed by Ness et al. (2011) and are common to more
 than one nova.
 
The X-ray light curve tell us about the resuming of accretion, the WD pulsations and underlying phenomena,
 the WD rotation, inhomegenities in the atmosphere, and sometimes about mass loss itself.

  The WD spectra were the most awaited, and they are challenging to analyse for the
 difficulties underlined in the previous section. 
 However, Fig. 4 and 5 clearly show that we have come a long way.
 The theories can be tested by assuming that T$_{\rm eff}$ is basically
 a proxy for  m$_{\rm WD}$, and the correlation
 of  m$_{\rm WD}$ with optical speed class and turn-off time (as indication of the accreted envelope
 mass) is the one expected.

  I would like to conclude 
 with an analysis of how we have started to answer the questions in Section 2. 

 1. So far, X-ray gratings' absorption spectra have shown signatures of a NeO WD only for V2491
 Cyg (Pinto et al. 2012).  The possibility of detecting these signatures is very interesting,
 because enhanced Ne, O or Mg in the ejecta may be due to mixing with traces of these elements
 in the WD superior layers, or with the secondary. The ejecta composition is not 
 necessarily the same as the underlying WD core, although mixing with the core occurs.
 It is important to probe the WD abundances directly.
 There is another interesting possibility to indirectly
 explore the WD composition, namely through nucleosynthesis
 of peculiar elements. Only mixing with Ne and Mg can produce  side reactions of the CNO cycle
 that synthesize Ar and Cl. Traces of these elements have been found in the optical
 spectra of some novae, and we detected transient Ar and Cl lines in a small soft flare that RS Oph
 showed at the very beginning of the SSS phase, in the {\sl XMM-Newton} observation of
 March 10 of 2006 (Orio et al. 2012, in preparation). Unfortunately, exact Ar and Cl overabundances
 in NeO novae depend on the uncertain 30P(p,g)  reaction (Iliadis et
 al. 1995). It would be important, after many years, to determine
 this reaction rate accurately (this is our special request from the nuclear physicists.

 2. For RS Oph and V4843 Sgr, the WD analysis through spectral elements shows a very large
 N/C ratio, typical of CNO ashes (Nelson et al. 2008, Rauch et al. 2010).
 This high ratio cannot be obtained if the nova is newly accreting or if
 its burning eroded WD material, so this is a proof of mass accretion over
 the secular evolution, despite the mass loss in the nova eruptions.

 3. The measured T$_{\rm eff}$ is consistent with the models' predictions, and
 Fig. 4 and 5 prove that the basic trends predicted by the models are verified.
 However, it would be important to explore a larger parameters' space,
 with more moderately slow and slow novae. Once a statistically significant number
 of novae are observed, the spread in linear T$_{\rm eff}$ versus turn-off
 and T$_{\rm eff}$ versus t$_2$ reilationships will indicate the range of variations
 of $\dot m$ and WD temperature and chemical composition at the onset of burning. 
 Thus, completing these plots is essential for the theories, and CCD-type instruments
 do not measur ethe absorption features, important for an accurate T$_{\rm eff}$ estimate.

 4. At least the hottest WD that we have observed with the gratings thanks to {\sl Swift}
 monitoring, must be very compact, with log(g) around 9. The luminosity would be well
 above the Eddington level otherwise.

 5. The emission line spectrum observed for RS Oph after
 other signatures of mass loss ceased, in June and September of 2006, seems to
 indicate is a residual wind for some time after the main eruption. 

   I would like to conclude with the predictions that acquiring more high resolution spectra
of novae in outburst will not only improve the statistics, but also throw new light on the
 data we already have.

\section*{Acknowledgements}
 The author wishes to acknowledge useful discussions with Thomas Rauch, Antonio
 Bianchini, Ehud Behar and Elia Leibowitz. Credit must be given to A. Bianchini
 for plotting Fig. 2.

\bigskip

\bigskip
\centerline{{\bf {\large References}}}

\noindent Balman, S., Krautter, J., \& \"Ogelman, H. 1998, ApJ, 499, 395\\
\noindent Balman, S., \& \"Ogelman, H. 1999, ApJ, 518, L111\\
\noindent Beardmore, A.P., et al. 2010, AN, 331, 156 \\
\noindent Becker, R.H., \& Marshall, F.E. 1981, ApJ, 244, L93 \\
\noindent Dobrotka, A., \& Ness, J.-U. 2010, MNRAS, 405, 2668 \\
\noindent Drake, J., et al. 2003, ApJ, 584, 448  \\
\noindent Eyres, S.P.S., et al. 2009, MNRAS, 395, 1533 \\
\noindent Heise, J., van Teeseling, A., Kahabka, P. 1994, A\&A, 288, L45 \\
\noindent Iliadis, C., Azuma, R.E., Buchmann, L., G\"orres, J., \& Wiescher, M. 1995, in 11Particle Physics and
 Cosmology: 9th lake Louise'', ARI \\
\noindent Kahabka, P. 1996, 306, 795 \\
\noindent Krautter, J., \"Ogelman, H., Starrfield, S., Wichmann, R., \&
Pfeffermann, E. 1996, \&A 456, 788  \\
\noindent Lanz, T., et al. 2005, ApJ, 619, 517 \\
\noindent Leibowitz, E., 2006, MNRAS, 371, 424 \\
\noindent Lloyd, H.M., et al. 1992, Nature, 356, 222 \\
\noindent Long, K.S., Helfand, D.J., Grabelsky, D.A. 1981, ApJ, 248, 925 \\
\noindent Nelson, T. et al. 2008, ApJ, 673, 1067 \\
\noindent Ness, J., et al.
\noindent Ness, J.-U., Starrfield, S., Jordan, C.,
Krautter J., \& Schmitt, J.H. M. M. 2005, MNRAS, 364, 1015
2003 \\
Ness, J.-U., et al. 2011, ApJ, 733, 70 \\
\noindent Ness, J.-U., et al. 2012, ApJ, 475, 43 \\
\noindent \"Ogelman, H., Orio, M., Krautter, J., \& Starrfield, S. 1993, Nature, 361, 331
Nelson, T., Orio, M. et al. 2008
\\
\noindent \"Ogelman, H., Krautter, J., \& Beuermann, K. 1984, ApJ, 287, L32
\\
\noindent \"Ogelman, H., Krautter, J., \& Beuermann, K. 197, A\&A, 177, 110
\\
\noindent \"Ogelman, H., Orio, M., Krautter, J., \& Starrfield, S. 1993, Nature, 361, 331
\\
\noindent Orio, M., Balman, S., della Valle, M., Gallagher, J., \& \"Ogelman,
 H. 1995, ApJ, 466, 410
\\
\noindent Orio, M., della Valle, M., Massone, G., \& "Ogelman, H. 1994, A\&A, 289, L11
\\
\noindent Orio, M. \& Greiner, J. 1999, A\&A, 344, L13
\\
Orio, M., Mukai, K., Bianchini, A., de Martino, D. \& Howell, S. 2009, ApJ, 690, 1573 \\
Orio, M. Hartmann, W., Still, M. \& Greiner, J. 2003, ApJ, 594, 435 \\
\noindent Orio, M., \"Ogelman, H., Covington, J. 2001, A\&A 373, 542 \\
\noindent Orio, M., et al. 2002, MNRAS, 333, L11 \\
\noindent Orio, M. et al. 2012, submitted \\
\noindent Orio, M. et al. 2012b, in preparation\\
\noindent Page, K.L., et al. 2009, A\&A, 527, 923
\\
\noindent Pinto, C., et al. 2012, A\&A, 543, 134 \\
\noindent Rauch, T. et al. 2010, ApJ. 717, 363 \\
\noindent Rohrbach, J. G., Ness, J.U., \& Starrfield, S. 2009, AJ, 137, 4627
\\
\noindent Sala, G., Hernanz, M., Ferri, M., \& Greiner, J. 2008, A\&A, 675, L93
\noindent Schwarz, G., et al. 2011, ApJS, 197, 31 \\
\noindent Shanley, L., \"Ogelman, L., Gallagher, J.S., Orio, M., \& Krautter, J. 1995, ApJ,
438, L95
\\
\noindent Strope, R.J., Schaefer, B.E., \& Henden, A. 2010, AJ, 140, 34 
\\
\noindent Shara, M. M., Prialnik, D., \& Shaviv, G. 1977, A\&A, 61, 363 
\\
\noindent Takei, D. et al. 2009, AJ, 137, 4160 \\
\noindent Tofflemire, B., Orio, M. 2012, in preparation \\
\noindent  
van den Heuvel, E.P.J., Bhattacharya, D., Nomoto, K., \& Rappaport, S. A. 1992, A\&A, 292, 97
\\
\noindent
 Yoon, S.C., \& Langer, N. 2005, A\&A, 435, 967
\end{document}